# Evidence for pressure-induced node-pair annihilation in $Cd_3As_2$

Cheng Zhang[1,2,†], Jianping Sun[3,†], Fengliang Liu[1,2,4,†], Awadhesh Narayan[5,6,†], Nana Li[4], Xiang Yuan[1,2], Yanwen Liu[1,2], Jianhong Dai[3], Youwen Long[3,8], Yoshiya Uwatoko[7], Jian Shen[1,2], Stefano Sanvito[6], Wenge Yang[4,9,\*], Jinguang Cheng[3,\*], Faxian Xiu[1,2,\*]

[1] State Key Laboratory of Surface Physics and Department of Physics, Fudan University, Shanghai 200433, China

[2] Collaborative Innovation Center of Advanced Microstructures, Nanjing, 210093, China

[3] Beijing National Laboratory for Condensed Matter Physics and Institute of Physics, Chinese Academy of Sciences, Beijing 100190, China

[4] Center for High Pressure Science and Technology Advanced Research (HPSTAR), Shanghai 201203, China

[5] Department of Physics, University of Illinois at Urbana-Champaign, Urbana, Illinois, USA

[6] School of Physics and CRANN and AMBER, Trinity College, Dublin 2, Ireland

[7] Institute for Solid State Physics, University of Tokyo, Kashiwa, Chiba 277-8581, Japan

[8] Collaborative Innovation Center of Quantum Matter, Beijing 100190, China

[9] High Pressure Synergetic Consortium (HPSynC), Geophysical Laboratory, Carnegie Institution of Washington, Argonne, IL 60439, USA

[†] These authors contributed equally to this work.

[*] Correspondence and requests for materials should be addressed to F. X. (E-mail: Faxian@fudan.edu.cn), J. C. (E-mail: jgcheng@iphy.ac.cn) or W. Y. (E-mail: yangwg@hpstar.ac.cn)

**Abstract**

As an intermediate state in the topological phase diagram, Dirac semimetals are of particular interest as a platform for studying topological phase transitions under external modulations. Despite a growing theoretical interest in this topic, it remains a substantial challenge to experimentally tune the system across topological phase transitions. Here, we investigate the Fermi surface evolution of $Cd_3As_2$ under high pressure through magnetotransport. A sudden change in Berry phase occurs at 1.3 GPa along with the unanticipated shrinkage of the Fermi surface, which occurs well below the structure transition point (~2.5 GPa). High-pressure X-ray diffraction also reveals an anisotropic compression of the $Cd_3As_2$ lattice around a similar pressure. Corroborated by the first-principles calculations we show that an axial compression will shift the Dirac nodes towards the Brillouin zone center and eventually introduces a finite energy gap. The ability to tune the node position, a vital parameter of Dirac semimetals, can have dramatic impacts on the corresponding topological properties such as the Fermi arc surface states and the chiral anomaly. Our study demonstrates axial compression as an efficient approach for manipulating the band topology and exploring the critical phenomena near the topological phase transition in $Cd_3As_2$.

**Introduction**

Located at the boundary of the topological phase diagram, Dirac semimetals can serve as a parent phase and be driven into many other exotic states, like Weyl semimetals or topological insulators, by explicitly breaking certain symmetries.[1, 2, 3] Recently, there is a surging interest in the phase transition of topological semimetals, as studied theoretically.[1, 4, 5, 6, 7, 8, 9, 10] Many exotic physical phenomena, such as quantum criticality[4, 8], metal-insulator transitions[7], and evolution of Fermi arc states[9], have been predicted to emerge at the critical transition points of topological semimetals. Despite the theoretical progress, experimental studies of these topological phase transitions remain largely unexplored. The difficulty lies in finding a way to effectively tune the electronic states across the topological phase transition. Evidence of a phase transition from Dirac semimetal to a semiconductor has only been observed in the Dirac semimetal $Cd_3As_2$ under high magnetic field.[3, 11] The Berry phase extracted from the quantum oscillations shows an unusual dependence on the direction and strength of the magnetic field.[3] However, a complex interplay between different kinds of symmetry breaking will emerge under such a circumstance since the magnetic field deviating from the crystal rotational axis breaks both the $C_4$ and the time-reversal symmetry. Furthermore, the large Landé *g* factor (~40) in $Cd_3As_2$ also contributes to a strong spin-orbit coupling with magnetic field.[12] Hence, a neat and more straightforward method of inducing topological phase transition in experiments remains to be discovered.

As widely used in condensed matter physics and geoscience, the application of high pressure provides a clean and effective means to modulate the physical and chemical properties of materials.[13] By directly tuning the lattice constant, high pressure can drastically impact the electronic structure even for strongly-correlated materials with high carrier density, something that is hard to achieve with other methods, for instance, with electrostatic gating. Many novel physical phenomena, such as quantum criticality and unconventional superconductivity, have been intensively studied with the application of high pressure.[14, 15] Unfortunately, several topological semimetals, including NbAs[16, 17], TaAs[18], and NbP[19], were reported to behave as "hard" crystalline lattice, where the external pressure over a moderate range cannot easily tune the Fermi surface (FS) and the band structure. Meanwhile, the complicated band structures in these materials make the detailed analysis on FS through the Shubnikov-de Haas (SdH) oscillations rather difficult. In contrast, the Dirac semimetal $Cd_3As_2$ has been shown to exhibit remarkable changes in transport properties by applying hydrostatic pressure or local stress.[20, 21, 22, 23] A structural phase transition from a metallic tetragonal phase to a semiconducting monoclinic one at around 2.57 GPa has been found in $Cd_3As_2$.[20] With the pressure further increased to 8.5 GPa, a superconducting transition emerges with $T_c$ around 2 K.[23] Similarly, the signature of unconventional superconductivity has also been reported by point-contact-induced local stress.[21, 22] These studies prove the validity of using high pressure as a tuning tool of the crystal and the band structure in $Cd_3As_2$. Nevertheless, it is still unclear how the FS evolves under external pressure, a crucial aspect for achieving a better understanding of these intriguing phenomena.

Here, we report a systematic magnetotransport study of $Cd_3As_2$ under hydrostatic pressures. Well-defined SdH oscillations are observed within the tetragonal phase below ~2.5 GPa and can be used to study the FS. Surprisingly, the FS cross-section shows an anomalous decreasing trend with increasing pressure. By analyzing the Landau fan diagram we uncover, in addition to the monoclinic structure transition, a hidden phase transition characterized by the sudden change in the phase factor of the SdH oscillations within the tetragonal phase. The jump of phase factor can be regarded as a direct consequence of an emerging band gap. In order to confirm this hypothesis, we perform a high-pressure X-ray diffraction (XRD) experiment on $Cd_3As_2$ crystals. An unusual increase in the *c/a* ratio around 1 GPa suggests an anisotropic compression behavior. By using advanced electronic structure calculations, we find that the unconventional response of the lattice constant to the external pressure shifts the Dirac nodes towards the Brillouin zone center and will eventually introduce a finite energy gap, therefore achieving a pressure-driven topological phase transition.

**Results**

**Transport across the tetragonal-to-monoclinic structure transition.** The transport measurements were performed on the (112) plane of the $Cd_3As_2$ single crystals (refer to Supplementary Fig. 1 for material characterizations). Figure 1a shows the temperature-dependent resistivity (*R-T*) curves of sample T1 under different pressures *P*. The resistivity of $Cd_3As_2$ increases with pressure in 1.0~7.5 GPa range. A clear transition in the slope of the *R-T* curve takes place at 2.5 GPa, corresponding to the structural transition from a tetragonal phase to a monoclinic one as reported earlier[20]. Such a clear difference can also be found in the magnetoresistivity (MR) ratio as shown in Fig. 1b. The MR ratio at 8.5 T decreases by nearly two orders of magnitude after the transition (see the inset in Fig. 1b). The Hall resistivity $\rho_{xy}$ at 2 K (Fig. 1c) exhibits a linear dependence on the magnetic field **B** both before and after the structural transition, suggesting that one type of carrier dominates the transport. By analyzing the Hall effect, we can determine the pressure dependence of the carrier density and the mobility as plotted in Fig. 1d. The carrier density abruptly increases from $2.5\times10^{18}$ cm$^{-3}$ to $1.5\times10^{19}$ cm$^{-3}$ at 2.5 GPa, followed by a continuous decrease at higher pressure. The high carrier density explains the metallic-like *R-T* curves in the high-pressure semiconducting phase reported elsewhere[20]. The electron mobility is reduced by over one order of magnitude after the transition, consistent with the previous report[20]. Accordingly, the quantum oscillations disappear immediately after the transition due to the relatively low mobility. For comparison, transport results at ambient pressure for another sample T3 from the sample batch of growth are presented in Supplementary Fig. 2.

**Pressure dependence of Fermi surface.** In order to elucidate how the band structure responds to the external pressure (*P*) within the tetragonal phase, we have systematically measured the magnetotransport of sample T2 in the metallic phase (0~2.17 GPa). Similarly to sample T1, the resistivity increases with pressure (Supplementary Fig. 3). The pure oscillations have been analyzed by subtracting a polynomial background from the MR. The extracted SdH oscillations are plotted as a function of 1/***B*** and *P* in Fig. 2a. The peaks and valleys of the oscillations are periodic in 1/***B*** and can be clearly tracked at different pressures. By employing fast Fourier transform (FFT) analysis, we can determine the oscillation frequency *F*. The frequency of SdH oscillations corresponds to a closed cyclotron orbit in *k*-space, described by $F = (\phi_0/2\pi^2)S_F$ with $\phi_0 = h/2e$. Here $S_F$ is the extremal cross-section area of the FS. Figure 2b summarizes the FFT spectra of the SdH oscillations at different $P$. The persistent single peak feature at different pressures indicates that only one cyclotron orbit (or a few degenerate ones) contributes to the oscillations, similarly to the ambient pressure case previously reported.[3, 24] A distinct behavior is that the oscillation frequency becomes smaller towards higher pressure. Conventionally, the FS tends to be larger upon compression due to the decrease of gap or the overlapping of the bands.[25, 26, 27] However, the anomalous decrease of FS with pressure suggests that the band structure experiences some changes other than the pure effect arising from the lattice shrinkage. Previous studies have shown that a Lifshitz transition takes place at around $E_{Lif} \approx 20$ meV above the Dirac points (corresponding to $k_{Lif} \approx 0.003$ Å$^{-1}$).[28] By assuming an isotropic three-dimensional FS (which is indeed the case for Cd$_3$As$_2$ at ambient pressure), we can calculate the Fermi vector at different pressures as shown in Fig. 2c. This clearly shows that the Fermi level of Cd$_3$As$_2$ is still well above the Lifshitz transition energy up to 2.17 GPa.

**A sudden jump in the phase factor.** The Berry phase is a phase difference acquired over a closed path in the parameter space during a cyclic adiabatic process.[29] Cyclotron orbits that enclose a Dirac point have been found to result in a nontrivial $\pi$ Berry phase.[29, 30] The Berry phase of a cyclotron orbit can be, in principle, detected from the SdH oscillations through the analysis of the phase factor.[30, 31] According to the Lifshitz-Onsager quantization rule $S_F \frac{\hbar}{eB} = 2\pi(n - \delta + \gamma)$, the phase factor $\gamma$, as part of the intercept in the Landau fan diagram, gives the Berry phase $\phi_B$ through the relation $\gamma = \frac{1}{2} - \frac{\phi_B}{2\pi}$, where $\delta$ (-0.125~0.125) is an additional phase shift originating from the curvature of the FS, $\hbar$ is the reduced Planck's constant and *e* is the electron charge. A nontrivial π Berry phase will lead to a phase factor near 0. As shown in Fig. 3a, we have performed a linear fit to the peak and valley positions as a function of 1/***B*** as integer and half-integer, respectively. There is a clear trend of the y-axis offset against external pressure in the Landau fan diagram (the inset of Fig. 3a and Fig. 3b). The intercept experiences a sudden jump around 1.3 GPa from ~0.13 at lower pressure to ~0.42 at higher pressure up to 2.17 GPa.

The change of intercept $\gamma - \delta$ could have several possible origins. At first sight, one may assume that the dimensionality of the FS changes under pressure. But the phase shift $\delta$ induced by the curvature of the FS is 0 for a 2D cylindrical FS and ±0.125 for a corrugated 3D FS (the – sign is for electrons, while + is for holes), with the precise value determined by the degree of dimensionality.[32,33] Since $Cd_3As_2$ remains *n*-type at all pressures, $\delta$ can only change from -0.125 to 0 at most. As such a change in the FS dimensionality cannot account for the observed phase shift. Hence, an alternative conjecture is that the phase factor $\gamma$ has changed.

In principle, the Berry phase should be quantized as 0 for normal fermions and π for Dirac fermions.[34] However, for massive Dirac fermions the dispersion is nonlinear at low energies, and the Berry phase is no longer path independent, but rather a function of the Fermi energy.[34] In this case a pseudospin orbital magnetic moment contribution should be taken into account in the cyclotron orbits. In an imperfect Dirac system with high-order corrections in the dispersion relationship (refer to Supplementary Fig. 4 for illustration), if a Dirac gap is generated, the Berry phase will change continuously from π to 0, depending on the gap size and the Fermi energy.[34, 35] Such a finite gap effect should be enhanced in strongly particle-hole asymmetric systems, like $Cd_3As_2$. In order to induce a gap into a Weyl system, one can only annihilate the Weyl node pairs with opposite chirality or break charge conservation through superconductivity.[36] The latter circumstance obviously does not happen in our experiment. As we know, the Dirac points in $Cd_3As_2$ are degenerate Weyl nodes protected by the $C_4$ symmetry.[37] Thus one way to destroy the Dirac points is by breaking the $C_4$ symmetry. In order to violate the $C_4$ symmetry, one could either apply a non-hydrostatic pressure or induce an intrinsic structure transition under isotropic pressure. Nevertheless, the high-pressure transport experiment of sample T2 was performed in a self-clamped BeCu piston-cylinder cell, which is a widely accepted method for achieving good hydrostatic pressure. Another possible way is to shift the two Dirac nodes and making them annihilate each other (two pairs of Weyl nodes), since here the system does not meet the requirement of the double Dirac semimetal phase.[38] Such scenario has been proposed in $Cd_3As_{2-x}P_x$ system where the inverted gap is generally eliminated by alloying with a light element to reduce the spin-orbit coupling.[39] So we can assume that a structure transition that breaks $C_4$ symmetry or a pressure-driven Dirac node shift leads to the observed gap opening behavior. Notably, the anomalous increase of resistance with pressure within the tetragonal structure reported previously[20] also supports the scenario of a band gap opening.

**XRD study of the crystal structure under high pressure.** In order to figure out how pressure may affect the band structure, we carried out high-pressure synchrotron XRD experiments with a symmetric diamond anvil cell. Figure 4a presents the angle-dispersive XRD patterns of $Cd_3As_2$ under different pressures. Similar to previous reports[20], the XRD spectrum experiences a significant change at 2.92 GPa, corresponding to the tetragonal to monoclinic structure transition. In the pressure range of 0.01~0.92 GPa, the XRD patterns show no significant changes with peaks systematically moving towards the higher angles. We performed Rietveld refinement on the XRD patterns at the low-pressure tetragonal phase to unveil the evolution of the tetragonal lattice structure in $Cd_3As_2$ under pressure. The obtained lattice parameters as a function of pressure are plotted in Fig. 4b-d. Both the unit-cell volume (Fig. 4b) and the lattice constants (Fig. 4c) decrease with increasing pressure. However, at around 1 GPa, the lattice constant $c$ experiences an abrupt contraction, which is absent in the volume and the lattice constant $a$. Considering the fact that the XRD results can be still well fitted by the tetragonal structure, the transition here should be an isostructural transition rather than a strong transformation between different point groups. Thus the $C_4$ symmetry is still preserved. The $c/a$ ratio shown in Fig. 4d further reveals this anomalous transition. As it can be seen, the $c/a$ ratio remains nearly unchanged in the regimes of 0.01~0.62 GPa and 1.24~2.52 GPa, but it presents a step-like feature at around 1 GPa. The $Cd_3As_2$ crystal exhibits different changes along $a(b)$ direction and $c$ direction, suggesting an anisotropic compression behavior.

**Theoretical calculations of the induced gap opening.** To study the influence of the anisotropic compression on the band structure, we carried out first-principles calculations within density functional theory (see Computational Methods for details). We considered several values of the *c/a* ratio, starting from *c/a*=2.01, which is the ambient pressure value, going down to *c/a*=1.95. The band structures for the different cases are shown in Fig. 5 a-d. As expected, for *c/a*=2.01, there is a Dirac crossing along the Γ-Z direction, which closely matches previous reports.[37] With decreasing *c/a* we notice a rather dramatic effect. Initially the Dirac nodes shift closer to the Brillouin zone center, becoming nearly overlapping for *c/a*=1.97. On further lowering *c/a* to 1.95, the time reversed Dirac partners annihilate and a band gap around 20 meV opens up (see Fig. 5e for a zoom close to zero energy). Figure 5f summarized the trend of energy gap and Lifshitz energy from the calculated band structure when decreasing *c/a* ratio. A systematic transition of the band structure from an inverted band to a normal one will take place as *c/a* decreases. In order to further support the evolution of band structure given by our *ab-initio* calculations, a low-energy model with anisotropic compression has been provided in Supplementary Note 1 by introducing a mass term $m_0$ into the effective Hamiltonian. As $m_0$ subsequently changes sign and becomes positive, a band gap is created. The effect of anisotropic compression is precisely that of reducing the mass inversion strength and can be mimicked by tuning the $m_0$ parameter in the low energy model. Remarkably, this is similar to the effect of alloying $Cd_3As_2$ with a lighter element, which also causes such a motion and annihilation of Dirac points.[39] Despite the slight difference of the critical *c/a* value given by theory and experiments, we found that their overall trend of energy gap when changing *c/a* is quite similar. Moreover, the anomalous shrinkage of the FS at high pressure (Fig. 2 b and c) observed in experiments is also consistent with the calculated band structure. Note that here we only focus on the effect of the *c/a* ratio on the band structure since it takes place at a pressure similar to that where the Berry phase in the transport suddenly changes. For comparison, another set of calculations by assuming isotropic compression was also performed as shown in Supplementary Fig. 5. The lattice constants were adopted from experimental value but the *c/a* ratio was fixed. In this scenario, the inverted gap and the Fermi vector are almost unchanged, which is incompatible with the transport results. Therefore, we attribute the anomaly in Fermi surface and Berry phase in $Cd_3As_2$ under high pressure to an anisotropic-compression-induced annihilation of the Dirac node-pair.

**Discussions**

Apart from the node-pair annihilation discussed above, there are other mechanisms that may lead to a change of the phase factor. First of all, a magnetic-field-induced phase shift has been observed in this particular system recently.[3, 11] It was found that high magnetic field applied in a direction, which deviates from the four-fold rotational axis, will make the Berry phase trivial.[11] The unique characteristic of this effect is that the Berry phase changes gradually with the increasing ***B***, while the field along the rotational axis will have no such effect. Here in this study, the magnetic field is perpendicular to the (112) plane of the sample. As reported earlier[3], the field along this direction will have a significant impact only when it is larger than 7 T. In the fitting process of the Landau fan diagram (Fig. 3a), most of the points are from the low field regime (below 6 T). At the same time, we have compared the phase factor extracted from the low field regime (refer to Supplementary Fig. 6). No such field dependence of phase factor is observed. Thus, we can safely exclude the magnetic field as the origin of the phase shift.

Another possible reason could be an anomalous phase shift which emerges near the Lifshitz point as proposed by Wang *et al.* recently.[40] In either the linear or parabolic limit, the energy spectrum is a simple function of $k_z^2$ with $\pm 0.125$ phase shift compared with that in 2D. Away from these two limits, this simple $k_z^2$ dependence is violated and it leads to a non-monotonic change in the phase shift.[40] Near the Lifshitz transition point, with a considerable component of the parabolic dispersion, this phase shift could be relatively large.[40] Meanwhile, additional beating pattern may be observed due to the Landau level crossing at extreme points of the band structure.[40] The major precondition for this anomalous phase shift is the mixture of linear and parabolic dispersions. As shown in Fig. 2c, the Fermi vector given by the SdH oscillations suggests that the Fermi level is well above the Lifshitz energy (~20 meV) given by our calculations and also previous experiments[12], more than ten times higher. And the Lifshitz energy further decreases with c/a ratio (refer to Fig. 5f). Moreover, the extracted SdH oscillations show clear single frequency, with no sign of beating pattern (refer to Supplementary Fig. 7). Hence, although we do not fully deny its existence, the present experimental evidences do not support this anomalous phase shift originating from the extreme point in band structure as a dominant contribution. Furthermore, the single-band oscillations also rule out other scenarios like the Zeeman splitting or the influence of a second band.

To conclude, we have reported a detailed study of $Cd_3As_2$ under high pressure through magnetotransport and XRD. Apart from the monoclinic structure transition, another hidden topological phase transition around 1.3 GPa is unveiled by the Fermi surface and Berry phase shift in the SdH oscillations. Moreover, an anomalous anisotropic compression also happens near the similar pressure range in the XRD experiment, which according to our calculations will annihilate the Dirac node pair and induce a finite energy gap. This work demonstrates that the pressurized $Cd_3As_2$ can be an ideal platform for further study of critical phenomena near the topological phase transition points.

**Methods**

**$Cd_3As_2$ single crystal growth**

$Cd_3As_2$ single crystals were synthesized by self-flux growth method in a tube furnace with stoichiometric amounts of high-purity Cd powder (4N) and As powder (5N) as reported elsewhere.[3] Mixed elements were sealed in an alumina crucible inside an iron crucible. The iron crucible was heated to 850℃, kept for 24 hours and slowly cooled down to 450℃ at 6℃/hour. Then the crucible was kept at 450℃ for more than one day before cooling. The superfluous Cd flux was removed by centrifuging after re-heated up to 450℃.

**High-pressure magneto-transport measurements**

High-pressure magneto-transport measurements have been performed under hydrostatic pressure up to 7.5 GPa (sample T1) with the cubic anvil cell (CAC) and up to ~ 2.17 GPa (sample T2) with a self-clamped BeCu piston-cylinder cell (PCC), respectively, in the Institute of Physics, Chinese Academy of Sciences. The pressure value in the CAC was calibrated at room temperature by the characteristic transitions of Bismuth (Bi) while that of PCC was determined by the superconducting transition temperature of Pb. The samples were shaped in to a rectangle with typical sizes of ~1.6 × 0.3 × 0.15 $mm^3$. Glycerol was chosen as the pressure transmitting medium.

**High-pressure synchrotron XRD study**

In-situ high-pressure synchrotron XRD experiment was performed under hydrostatic pressure in the symmetric diamond anvil cell with the diamond anvil culet size of 300 μm. Firstly, a Rhenium gasket was indented to about 15 GPa, resulting in a 50 μm gasket thickness. Then a hole with the diameter of 150 μm was drilled in the center by the laser, serving as the sample chamber. A small piece of $Cd_3As_2$ sample was put into the chamber, together with two small ruby balls for pressure calibration[41] and silicone oil as pressure medium. The XRD experiment was performed at 13-BM-C, GSECARS, APS, Argonne National Laboratory, with the incident X-ray wavelength of 0.434 Å. The original XRD pattern was integrated by the Fit2D software, followed by the Rietveld refinements through the GSAS software to extract the lattice parameters.

**Computational methods**

Density functional theory calculations were carried out using the Vienna Ab-initio Simulation Package (VASP).[42] Perdew-Burke-Ernzerhof form of the exchange-correlation functional was used.[43] A plane wave cut-of 300 eV was employed and Brillouin zone was sampled using a $4\times4\times2$ $k$-point grid. Spin-orbit coupling was included in all calculations. Experimentally obtained lattice parameters were used.

**Acknowledgements**

F.X. was supported by the National Young 1000 Talent Plan, National Natural Science Foundation of China (61322407, 11474058, 61674040). Part of the sample fabrication was performed at Fudan Nano-fabrication Laboratory. J.C. acknowledges the support of the MOST and NSF of China (Grant Nos. 2014CB921500, 11574377, and 11304371), the Strategic Priority Research Program and the Key Research Program of Frontier Sciences of the Chinese Academy of Sciences (Grant No. XDB07020100 and QYZDB-SSW-SLH013), and the Opening Project of Wuhan National High Magnetic Field Center (Grant No. 2015KF22), Huazhong University of Science and Technology. SS thanks the European Research Council (QUEST project) for financial support. W.Y. was supported by NSAF (Grant No U1530402). We acknowledge Dr. Sheng Jiang of 15U1 at SSRF, Dr. Dongzhou Zhang of 13-BMC at GSECARS, APS, and Dr. Daijo Ikuta of 16-BMD at HPCAT, APS for technical support. HPCAT operations are supported by DOE-NNSA under Award No. DE-NA0001974 and DOE-BES under Award No. DE-FG02-99ER45775, with partial instrumentation funding by NSF. 13BM-C operation is supported by COMPRES through the Partnership for Extreme Crystallography (PX2) project, under NSF Cooperative Agreement EAR 11-57758. APS is supported by DOE-BES, under Contract No. DE-AC02-06CH11357. Computational resources were provided by the Trinity Centre for High Performance Computing.

**Author contributions**

F.X. conceived the ideas and supervised the overall research. C.Z. Y.L. and X.Y. synthesized $Cd_3As_2$ single crystal. J.S. and J.C. carried out the magneto-transport measurements under high pressures with the help from J.D., Y.W.L. and Y.U. C.Z. analysed the transport data. F.L. N.L. J.S. and W.Y. performed the high-pressure crystal structure study. A.N. and S.S. provided the theoretical model. C.Z. and F.X. wrote the paper with helps from all other co-authors.

## Competing financial interests

The authors declare no competing financial interests.

# FIGURES

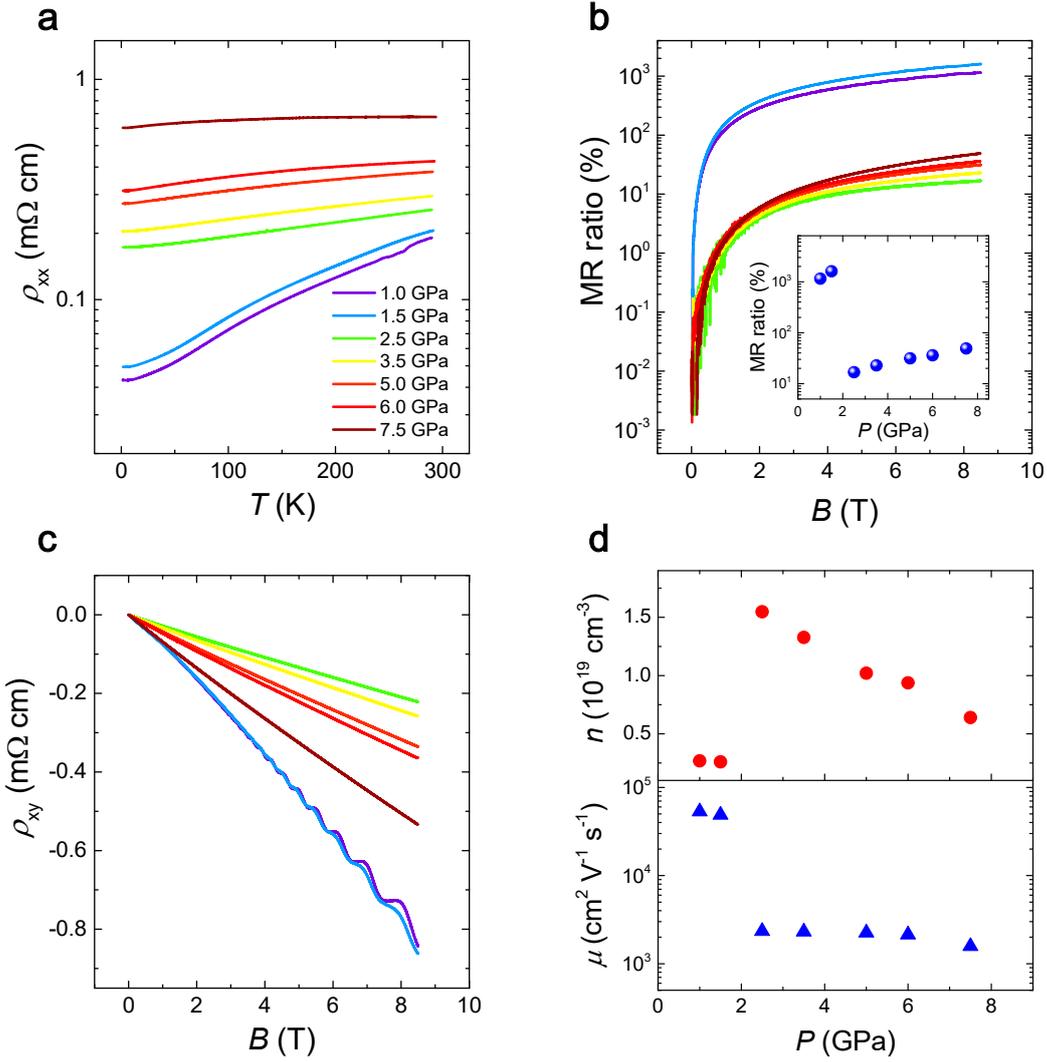

**Figure 1| High pressure transport measurements across the structure transition in sample T1.**
(**a**) The temperature dependence of resistivity (*R-T*) curve under different pressures. (**b-c**) The magnetic field dependence of MR ratio (**b**) and Hall resistivity (**c**) under different pressures at 2 K. The inset of **b** is the MR ratio at 8.5 T. (**d**) The carrier density *n* and mobility *μ* at 2 K under different pressures.

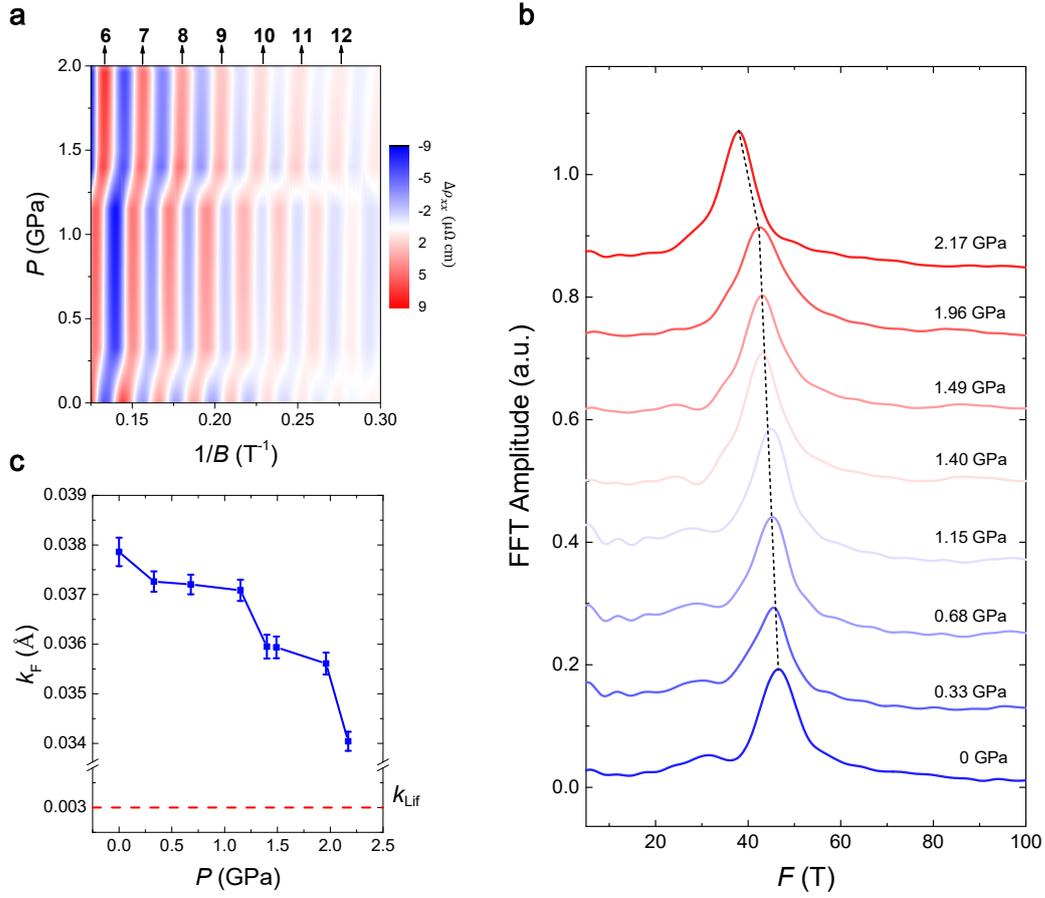

**Figure 2| The pressure dependence of SdH oscillations in sample T2.** (**a**) The extracted SdH oscillations as a function of 1/$B$ and pressure at 2 K. The peaks are marked as a series of Landau level index. (**b**) A stack view of FFT spectrum of SdH oscillations at different pressures. The FFT peaks shift towards small value at higher pressures. (**c**) The pressure dependence of Fermi vector $k_F$. The corresponding Fermi vector for Lifshitz transition at ambient pressure is marked as $k_{Lif}$ indicated by the red dash line.

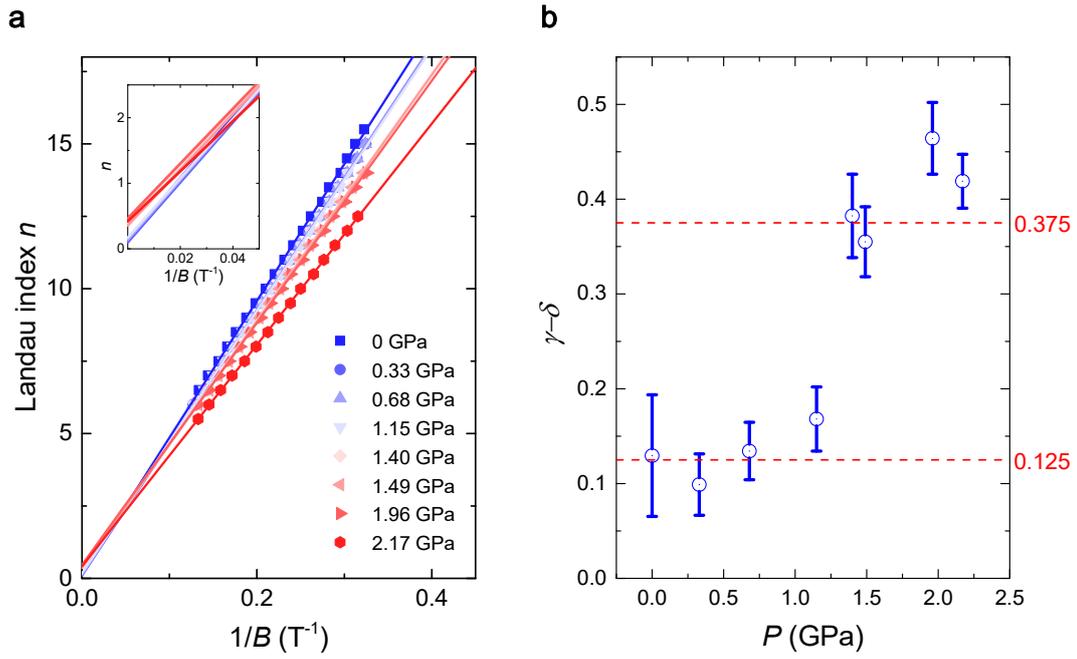

**Figure 3| The Landau fan diagram and phase factor at different pressures.** (**a**) The Landau fan diagram of sample T2 under different pressures. The inset of **a** is a close look of the y-axis intercept. (**b**) The extracted phase factor $\gamma - \delta$ under different pressures. A sudden jump occurs around 1.3 GPa.

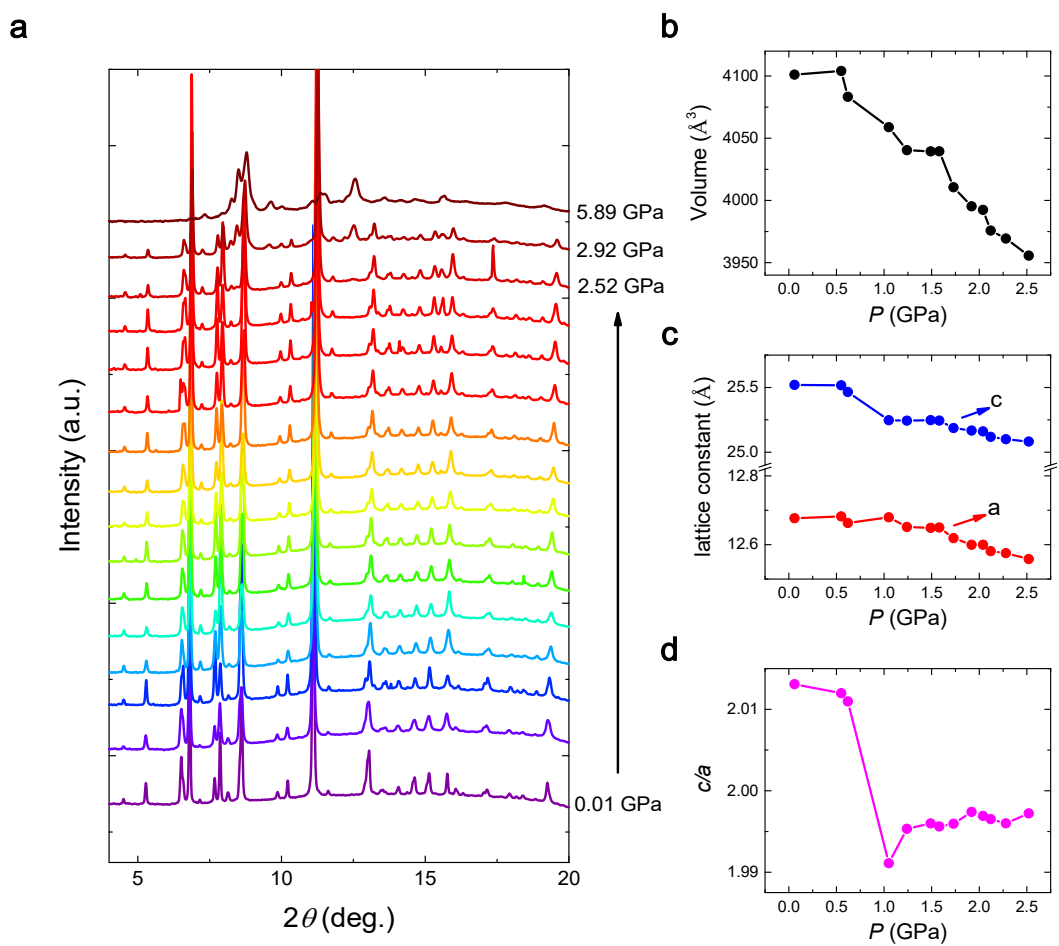

**Figure 4| The high-pressure XRD study of Cd$_3$As$_2$.** (**a**) A stack view of XRD spectrum under different pressures. The sudden change of spectrum at 2.52 GPa corresponds to a tetragonal-to-monoclinic structure transition. (**b**) The pressure dependence of unit cell volume, (**c**) the lattice constant *a* and *c*, and (**d**) the lattice constant ratio *c/a*.

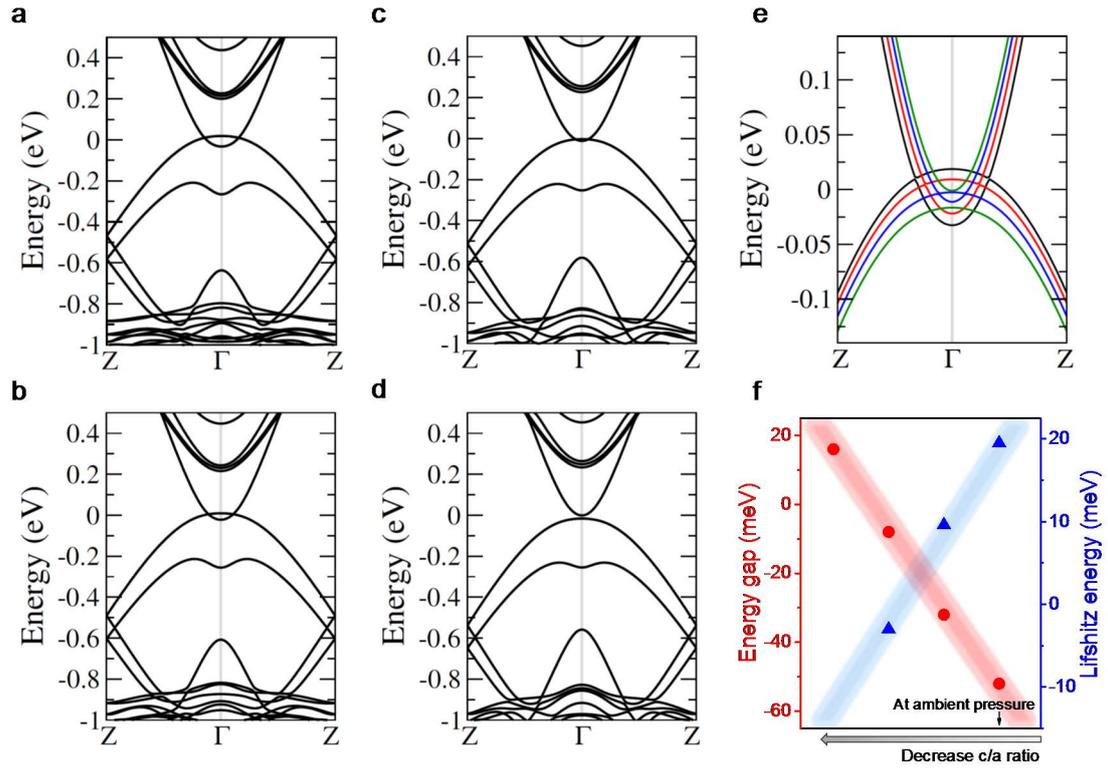

**Figure 5| First-principles band structure for $Cd_3As_2$ at different c/a ratios.** (**a-d**) The calculated band structure with the *c/a* ratio being 2.01 (**a**), 1.99 (**b**), 1.97 (**c**), and 1.95 (**d**), respectively. (**e**) Zoom around the Dirac points comparing the band structures at the different c/a ratios. Black, red, blue, and green curves correspond to **a-d**, respectively. (**f**) The evolution of energy gap at Γ point and Lifshitz energy with decreasing *c/a* ratio. The negative value of energy gap corresponds to an inverted band. The Lifshitz energy decreases with *c/a* ratio and a small gap around 20 meV emerges when *c/a* reaches 1.95. Note that the Lifshitz transition is absent when a positive energy gap is generated.